 \documentclass[preprint2]{aastex}

\slugcomment{{\it Science} 2006 {\bf 313}, 935}

\shorttitle{Pinwheels in the Quintuplet}
\shortauthors{Tuthill et al.}

\newcommand{\simle}{\mbox{$\stackrel{<}{_{\sim}}$}}

\begin{document}

\title{Pinwheels in the Quintuplet Cluster}

\author{Peter~Tuthill,\altaffilmark{1}
        John~Monnier,\altaffilmark{2}
        Angelle~Tanner,\altaffilmark{3}\\
        Donald~Figer,\altaffilmark{4}
        Andrea~Ghez,\altaffilmark{5}
\&    William~Danchi,\altaffilmark{6}}

\altaffiltext{1}{School of Physics, University of Sydney, NSW 2006, Australia} 
\altaffiltext{2}{Univ. Michigan Astronomy Dept., 500 Church Street, Ann Arbor, MI 48109-1090, USA} 
\altaffiltext{3}{JPL/Caltech, Mail Code 100-22, 770 South Wilson Ave, Pasadena, CA 91125 USA} 
\altaffiltext{4}{COS/Ctr for Imaging Sci, 54 Lomb Memorial Dr, RIT, Rochester, NY 146235604, USA} 
\altaffiltext{5}{Astronomy, UCLA, Los Angeles, CA 90095-1547, USA} 
\altaffiltext{6}{NASA Goddard, Code 667, Greenbelt, MD 20771, USA}

\email{Correspondance to Peter Tuthill: gekko@physics.usyd.edu.au}

\begin{abstract}
The five enigmatic ``Cocoon stars" after which the Quintuplet cluster was
christened have puzzled astronomers since their discovery. Their extraordinary 
cool, featureless thermal spectra have been attributed to various stellar types 
from young to highly evolved, while their absolute luminosities places them 
among the supergiants. We present diffraction-limited images from the Keck 1 
telescope which resolves this debate with the discovery of rotating spiral 
plumes characteristic of colliding-wind binary ``pinwheel" nebulae. 
Such elegant spiral structures, found around high-luminosity Wolf-Rayet stars, 
have recently been implicated in the behavior of supernovae lightcurves in the
radio and optical.
\clearpage
\end{abstract}

The five enigmatic ``Cocoon stars'' after which the Quintuplet cluster was christened
have puzzled astronomers since their discovery ({\it 8}).
Hundreds of stars have now been identified within the cluster ({\it 1, 2}),
placing it among the most massive in our Galaxy, yet the nature of the five extremely red
stars at the heart of the Quintuplet has remained elusive.
Their extraordinary cool, featureless thermal spectra ($\sim$780-1315\,K ({\it 1}))
have been attributed to various stellar types from young to highly evolved, 
while their absolute luminosities places them among the supergiants
($10^{4-5}$\,L$_\odot$).
We present diffraction-limited images from the Keck~1 telescope which resolves this 
debate with the discovery of rotating spiral plumes characteristic of colliding-wind 
binary ``pinwheel'' nebulae.
These have previously been reported in dust shells around luminous hot Wolf-Rayet stars
({\it 11, 6}).

Using high-resolution speckle techniques in the near-infrared ({\it 13}), 
all five Cocoon stars were (at least partially) resolved, and images recovered
of the two presenting the largest apparent size, Q\,2 and Q\,3, (Fig.~1).
Outflow plumes depicted follow the form of an Archimedian spiral, thereby establishing the 
presence of circumstellar dust formed in a colliding-wind binary system.
These rare Pinwheel nebulae result when dust condensation in the stellar wind is mediated 
by the presence of a companion star, as established in the prototype systems WR~104 ({\it 11})
and WR~98a ({\it 6}).
Dust nucleation is enabled by the wind compression associated with the bowshock between the 
stellar winds.
Newly-formed hot dust streaming into the wake behind the companion is wrapped into a 
spiral by the orbital motion as it is embedded within the expanding Wolf-Rayet wind.

With high-resolution images available from two epochs separated by 357\,days, the
dust plume of star Q\,3 was fitted to an Archimedian spiral model with winding angle
110$\pm$10\,mas/turn and with an inclination of the rotation axis to the line-of-sight 
of 26$^\circ$ (not well constrained; adequate fits possible in the range 0--36$^\circ$).
The proper motion of structures between the two epochs indicates a rotation
period of the spiral, and hence the colliding-wind binary, of 850$\pm$100\,days.
Assuming an 8000\,pc distance to the Quintuplet ({\it 9}), these measurements
constrain the Wolf-Rayet wind velocity to be $v_\infty$=1800$\pm$300\,km/sec, which is
typical for a late-type Carbon-rich (WC-spectrum) star ({\it 4}).

Simple models fall short of reproducing all structures (Fig.~1), 
particularly near the bright core where multiple knots and streams can exist. 
This has also been noted in WR~98a ({\it 6}) where extra complexity was attributed 
to optical depth and line-of-sight effects from a 3-D structure ({\it 3}).
Q\,2 appears to have similar parameters to Q\,3 although a second image epoch is required
to measure the rotation rate. 
Partially-resolved objects Q\,1,  Q\,4 and Q\,9 (see {\it 13} and Table~S1 for observational
results) exhibit similar colors and surface brightness to Q\,2 \& Q\,3, and we therefore suggest 
that they are also pinwheels but with tighter winding angles (therefore shorter periods), or 
less favorable inclinations.
Furthermore, the prototype pinwheel WR~104, with a period of 243\,days, would give an
apparent size in close accord with measured sizes of Q\,1,  Q\,4 and Q\,9 if it were 
removed to the distance of the Quintuplet.

Given the extreme visible extinction ($A_v$=29$\pm$5 ({\it 2})), small separation
of the central binary stars (likely $\sim$0.6\,mas or 5\,AU), and presence of high-luminosity 
circumstellar dust shells, it would be extremely difficult to detect or study these
systems with other techniques.
The most luminous stars in our Galaxy are often surrounded by dusty shells, 
and the implication that most, if not all of these harbor massive binaries
(not single stars) has important ramifications for the high-mass tail of the 
stellar Initial Mass Function (IMF).
Binarity is also a key element to studies of Type~Ib/c and Type~IIb supernovae. 
There are recent indications that explosion lightcurves can be modified by the imprint 
of circumstellar matter, carrying an encoding of the mass-loss history of the supernova
precursor star system ({\it 7, 10}).

\begin{quote}
{\bf References and Notes}

\begin{enumerate}

\item Figer, D.~F., McLean, I.~S., \& Morris, M.\ 
      1999, {\it ApJ}, {\bf 514}, 202

\item Figer, D.~F., Kim, S.~S., Morris, M., Serabyn, E., Rich, R.~M., \& McLean, I.~S.\ 
      1999, {\it ApJ}, {\bf 525}, 750 

\item Harries, T.~J., Monnier, J.~D., Symington, N.~H., \& Kurosawa, R.\ 
      2004, {\it MNRAS}, {\bf 350}, 565

\item Howarth, I.~D., \& Schmutz, W.\ 
      1992, {\it A\&A}, {\bf 261}, 503

\item  Moneti, A., Glass, I.~S., \& Moorwood, A.~F.~M.\ 
      1994, {\it MNRAS}, {\bf 268}, 194

\item Monnier, J.~D., Tuthill, P.~G., \& Danchi, W.~C.\ 
      1999, {\it ApJ}, {\bf 525}, L97

\item Moran, J.~A., \& Reichart, D.~E.\ 
      2005, {\it ApJ}, {\bf 632}, 438

\item Okuda, H., et al.\ 
      1990, {\it ApJ}, {\bf 351}, 89

\item Reid, M.~J.\ 
      1993, {\it ARA\&A}, {\bf 31}, 345

\item Ryder, S.~D., Sadler, E.~M., Subrahmanyan, R., Weiler, K.~W., Panagia, N., \& Stockdale, C.\
      2004, {\it MNRAS}, {\bf 349}, 1093

\item Tuthill, P.~G., Monnier, J.~D., \& Danchi, W.~C.\ 
      1999, {\it Nature}, {\bf 398}, 487

\item We thank Stuart Ryder, Charles Townes, Eric Becklin, and Seth Hornstein for
      contributions to this paper.
      This work has been supported by the Australian Research Council and the
      U.S. National Science Foundation.
      Data presented here were obtained at the W.M. Keck Observatory.

\item Observational methods and discussion are available in {\it Science} online:\\
      www.sciencemag.org (contents: Observations and Discussion, Table~S1 and Fig.~S1)
      For Astro-PH, this material follows in the next section.
\end{enumerate}
\end{quote}


\begin{figure*}[B!]
\mbox{
\includegraphics[angle=90,height=7.2cm]{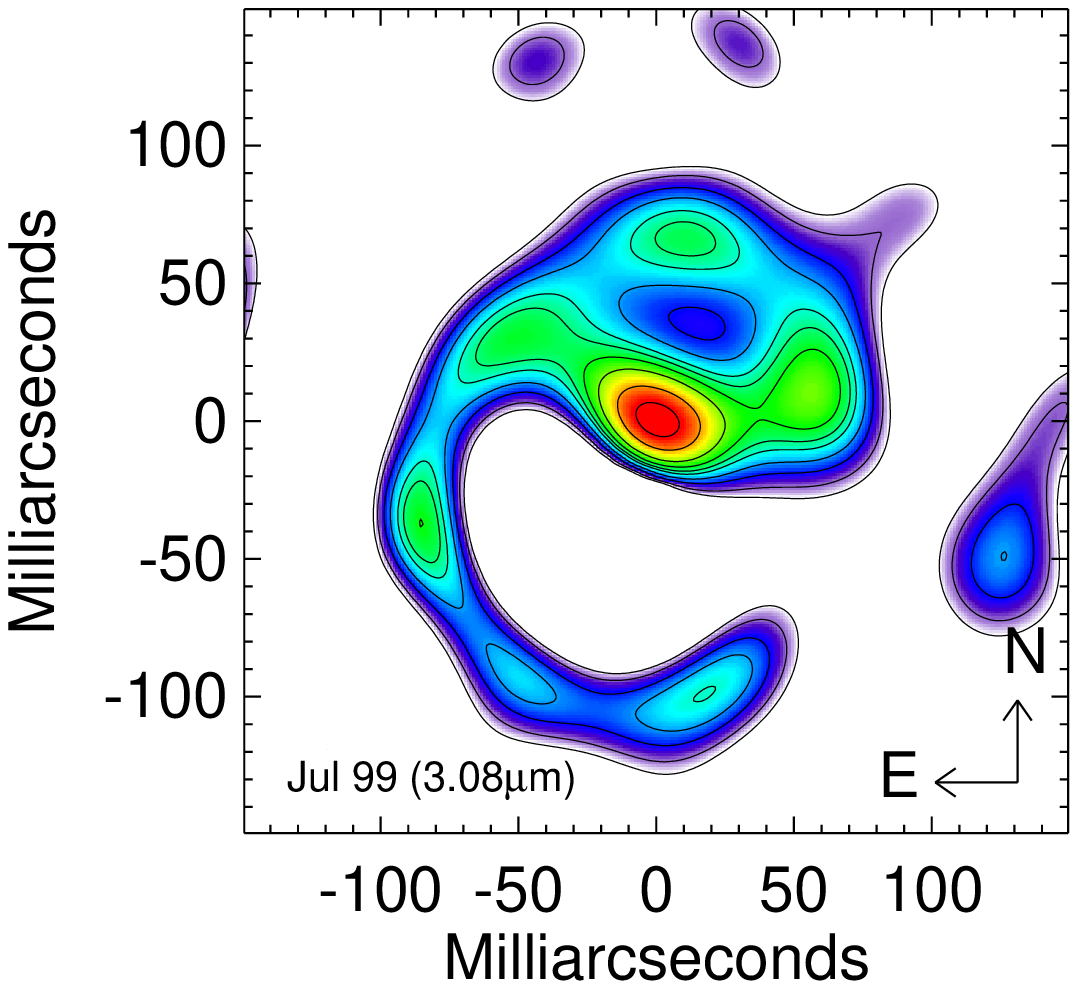}
\includegraphics[angle=0,width=10.0cm]{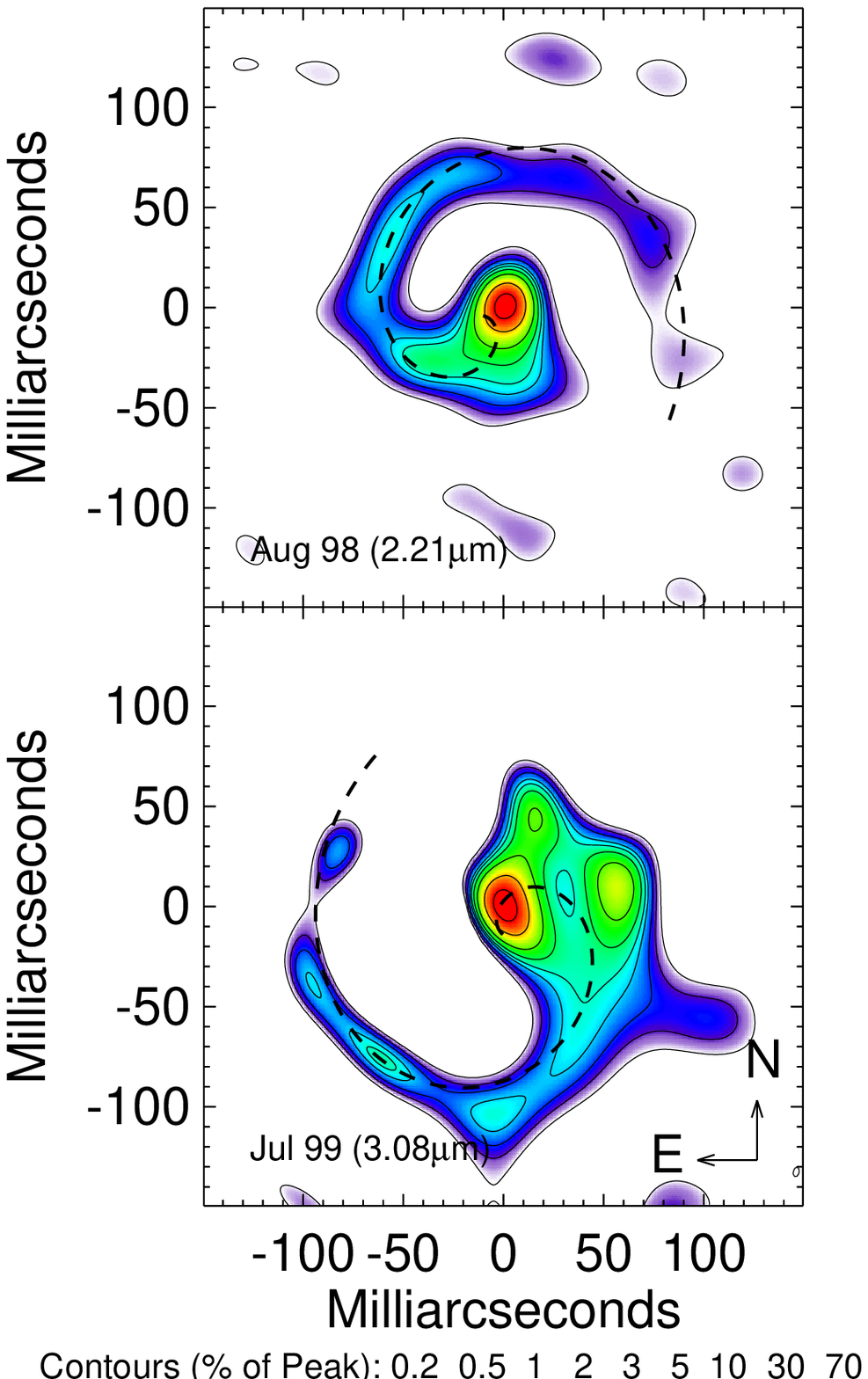}}
\caption{ False-color images of Q\,2 at 3.08\,$\mu$m 1999~July (left panel) 
and Q\,3 at 2.21\,$\mu$m from 1998~Aug (upper right panel) and 3.08\,$\mu$m from 
1999~July (lower right panel).
Overplotted on the Q\,3 images is a rotating Archimedian spiral model fitted 
to the dominant tail of the outflow plume at the two separate epochs (dashed line).
Identification of Quintuplet objects including Q\,2, Q\,3 is discussed further in 
the supporting online material and Fig.~S1.
\label{qfig}}
\end{figure*}

\clearpage

{\bf \Large Supporting On-line Material}

\section{Observations and Discussion}

Observations of the Quintuplet cluster are listed in Table~S1 and were made with 
the near-infrared NIRC camera on the Keck~1 telescope.
The nomenclature of Moneti et al.\citep{moneti94} is adopted here, identifying the dusty red 
Quintuplet proper members, or cocoon stars, as Q\,1, Q\,2, Q\,3, Q\,4 and Q\,9. 
Locations of each of these stars within the cluster as imaged by the Hubble Space Telescope
is given in Fig.~S1, which also depicts our diffraction-limited Keck images of Q\,2, Q\,3 
overlayed with graphical indication of the relative spatial scale.

The high angular resolution imaging utilized rapid-exposure speckle interferometry 
techniques from a number of long-standing experiments at the Keck~1 \citep{keckmask}.
Typical datasets entailed from one to a few hundred short-exposures ($\sim$0.14\,sec)
with the NIRC camera operating in a high-magnification (0.0206 arcsec/pixel) mode.
Interleaved observations of galactic-center object IRS~7 were used to calibrate the 
telescope-atmosphere point-spread function (the unresolved nature of this object was 
itself checked against stars HD~159255 and HD~163042).

All five Quintuplet cocoon stars were found to be spatially resolved to some degree.
The full-width at half-maximum (FWHM) of best-fitting circular Gaussian functions, used
here to give an estimate of apparent size, are given in Table~S1.
Images of stars Q\,2 and Q\,3, recovered with Fourier techniques\citep{keckmask} and 
presented earlier in Fig.~1, show that complex and asymmetric structures exist within 
the dust shells surrounding these stars.
The complexity of these visibility functions resulted in poor fits for the Gaussian 
model, and we therefore limited the fit range to low spatial frequencies/short baselines
where visibility excursions are not yet strongly manifest
($<$2$\times$10$^6$ and 1$\times$10$^6$\,rad$^{-1}$ for 2.21 and 3.08\,$\mu$m
respectively: equivalent to $\simle$4.5 and 3.1\,m).

By implication, if similar physics is assumed to pertain to the remaining three of the 
five Quintuplet cocoon stars, then the utility of fitting Gaussian shells may
appear limited.
Despite this, for objects which are only partially resolved, the fitting of such simple 
functional forms to interferometer data gives a good estimate of the overall size, and
permits quantitative comparison with models.
Furthermore, this allows sizes for morphologically complex objects Q\,2 and Q\,3
to be compared against measurements of the partially-resolved Q\,1, Q\,4 and Q\,9, 
and against other dusty Wolf-Rayets observed with high resolution techniques.
Note also that FWHM in the range 10--15\,mas are only marginally resolved 
with this experiment; relative errors in these cases are correspondingly high.

Apparent sizes have been combined with flux measurements to yield estimates of surface
brightness in our two filter bandpasses for the target stars.
Correction for the $A_v$=29$\pm$5 visible extinction was made using the optical dust constants
of Mathis\citep{mathis90}.
Surface brightnesses were then derived following Monnier et al.\citep{wr_size}, which also
gives a comparison population of Galactic dusty Wolf-Rayets with similarly measured surface 
brightnesses.
For both filter bandpasses, the Quintuplet cocoon stars showed surface brightnesses within
the range spanned by the WR population studied in Monnier et al.\citep{wr_size}.
In particular, for Q4 and Q9 the color temperature between 2.21 and 3.08\,$\mu$m appeared to 
be in reasonable accord with similar measurements from the Galactic population.
This finding was found to be generally robust against variations in the extinction correction
over the expected range.

However, in some regards the Quintuplet WRs, and in particular Q\,2 and Q\,3, did appear to
be distinct from the Galactic population.
The observed increase in size between 2.21 and 3.08\,$\mu$m which approaches a factor of $\sim$2
in the Quintuplet, is found to be a more modest $\sim$1.4 elsewhere\citep{wr104-idlm,wr_size}.
Such changes of size (and surface brightness) with wavelength reflect the fact that dust with
a range of temperatures contributes to the near-IR emission. 
The dramatic enlargement between 2.21 and 3.08\,$\mu$m argues for a flatter thermal profile
in the Quintuplet dust shells: plausibly due to external heating from stars in the dense central 
region of the cluster. 
The comparison population of Galactic WRs were in far less crowded regions where the outer 
dust shell likely receives little or no energy except that originating with the central WR
star.

Perhaps the simplest way to test the hypothesis that the remaining 3 cocoon stars 
are also pinwheels is by photometric monitoring.
Variability of these sources has already been noted\citep{glass02}, although not
with sufficient coverage to reveal a cyclic change consummate with any
rotational period\citep{kato02}. 
The $\sim$2.5\,yr periods inferred from the imaging extend the confirmed 
operation of continuous dust formation by the pinwheel mechanism to significantly 
longer periods/larger binary separations than previously known, with implications 
for models of these processes.

This finding of late-type WC binaries in the Quintuplet means that WC (Carbon rich)
outnumber WN (Nitrogen rich) Wolf-Rayets by 11:6, and furthermore all WC stars are dusty.
This makes an interesting contrast to the massive young WR-rich cluster Westerlund~1
where the WC:WN ratio is reversed to 7:12, with none of the WC's exhibiting dust.
Clearly, the close binaries at the heart of the Pinwheel systems can be
responsible for significant modification of the stellar evolutionary path, in which
mass-transfer or envelope-stripping events might precipitate the WC phase.
This entanglement of binarity, mass-loss history, and evolutionary path has the potential
to skew population distributions, although it is unclear exactly what conditions resulted
in the abundance of Pinwheels in this cluster.

\clearpage

\renewcommand{\thetable}{S1}
\begin{deluxetable}{clccc}
\tablewidth{0pt}
\tablecaption{\label{tbl-log}
Observing log and apparent sizes}
\tablehead{
\colhead{Source\cite{moneti94}} & \colhead{Alt. Name\cite{okuda90}} 
& \colhead{Date} &\colhead{2.21\,$\mu$m FWHM (mas)} & \colhead{3.08\,$\mu$m FWHM (mas)} }
\startdata
Q\,1 & GCS 3-4 & 1998 Aug 06 & 18$\pm$3 & -- \\
Q\,2 & GCS 3-2 & 1998 Aug 06 & 35$\pm$2 & -- \\
     &         & 1999 May 04 & 38$\pm$3 & -- \\
     &         & 1999 Jul 29 & 37$\pm$3 & 77$\pm$2 \\
Q\,3 & GCS 4   & 1998 Aug 06 & 41$\pm$2 & -- \\
     &         & 1999 Jul 29 & 40$\pm$2 & 75$\pm$2 \\
Q\,4 & GCS 3-1 & 1998 Aug 06 & 13$\pm$4 & -- \\
     &         & 2002 Jul 23 & 15$\pm$3 & 20$\pm$3 \\
Q\,9 & GCS 3-3 & 1998 Aug 06 & 13$\pm$4 & -- \\
     &         & 2002 Jul 23 & $<11$ & 21$\pm$3 \\
\enddata
\tablecomments{~Log of observations of the five cocoon stars.
Observations in the K band at 2.21\,$\mu$m were made at all
epochs, but longer wavelength data at 3.08\,$\mu$m were only
secured on the four occasions listed.  
The FWHM of a circular Gaussian profile fit to the visibility
data is also given, together with the estimated uncertainty,
as a measure of the overall apparent size.}
\end{deluxetable}

\clearpage

\begin{center}
\includegraphics[angle=0,width=14cm]{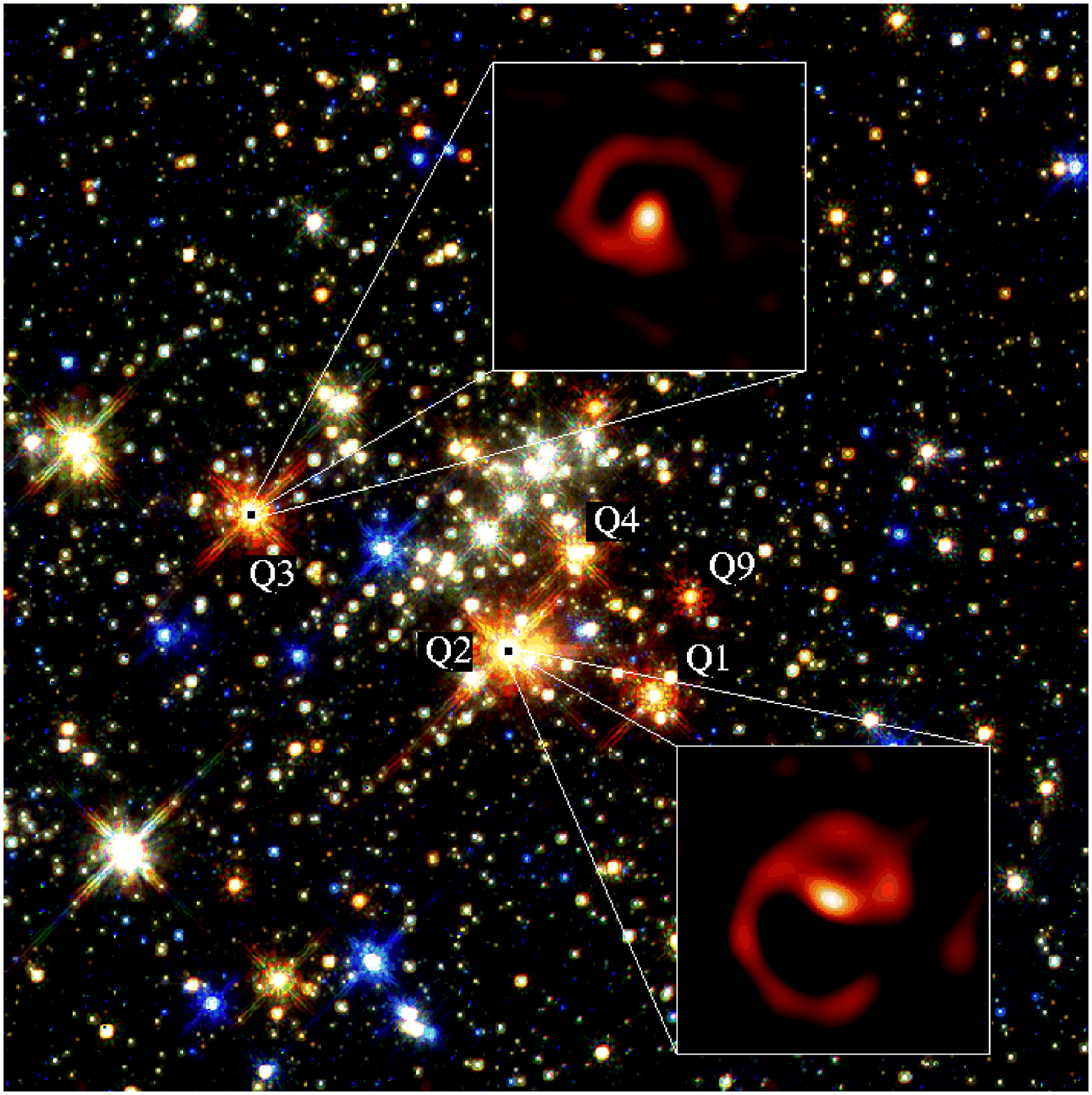}
\end{center}
{Fig.~S1-- The background star-field is from multi-wavelength Hubble Space Telescope 
(NICMOS) near-infrared imaging.
Further details and discussion of this image can be found in Figer et al. \cite{figer99b}.
The five dusty red cocoon stars are labelled according to the nomenclature of Moneti et al.
\citep{moneti94}.
Inset images of Q\,2 and Q\,3 recovered with our Keck imaging experiments (see also Fig.~1) 
are overlayed, with graphical indication showing the relative scaling between the Hubble and
Keck imaging. 
}

\end{document}